\documentclass{eas}
\usepackage{graphicx,amsmath,amsmath,epsf,pstricks,amsfonts,epsfig}
\begin{document}

\title{Quantum vacuum and accelerated expansion}
\runningtitle{Broda {\em et al.\/}: Quantum vacuum and accelerated
expansion}
\author{Bogus{\l}aw Broda}\address{Department of Theoretical Physics, University of
{\L}\'od\'z\\ Pomorska 149/153, 90--236 {\L}\'od\'z, Poland;\\
\email{bobroda@uni.lodz.pl\ \&\ michalszanecki@wp.pl}}
\author{Micha{\l} Szanecki}
\sameaddress{1}
\begin{abstract}
A new approach to extraction of quantum vacuum energy, in the
context of the accelerated expansion, is proposed, and it is shown
that experimentally realistic orders of values can be derived. The
idea has been implemented in the framework of the
Friedmann--Lema\^{\i}tre--Robertson--Walker geometry in the
language of the effective action in the relativistic formalism of
Schwinger's proper time and Seeley--DeWitt's heat kernel
expansion.
\end{abstract}
\maketitle
\section{Introduction}
The following three, well-known problems of modern physics and
cosmology, accelerated expansion of the Universe (Riess {\em et
al.\/} \cite{Riess1998}; Perlmutter {\em et al.\/}
\cite{Perlm1999}), very small but non-vanishing cosmological
constant or dark energy (Weinberg \cite{Wein1989}), (Carroll
\cite{Carr2001}; Padmanabhan (\cite{Padm2003}, \cite{Padm2006})),
and theoretically extraordinarily huge quantum vacuum energy
density (Zel`dovich \cite{Zeld1967}), (Volovik (\cite{Vol2005},
\cite{Vol2006})) can be treated as mutually related or as
independent problems. An old approach to the issue of the
cosmological constant $\Lambda$ utilizes quantum vacuum energy as
a solution of this issue, but unfortunately it does not work
properly. Namely, it appears that directly calculated,
Casimir-like value of quantum vacuum energy is more than one
hundred orders greater than expected. Such a huge value of quantum
vacuum energy is a serious theoretical problem in itself. Lowering
the UV cutoff scale down from the planckian to the supersymmetric
one is a symbolic improvement (roughly, it cuts the order by two
(Weinberg \cite{Wein1989})). A more radical reduction of the
cutoff could cure the situation but it would create new problems.
Sometimes, it is claimed that vacuum energy for one or another
reason does not influence gravitational field.

In this paper, following ideas presented in (Broda {\em et al.\/}
\cite{Broda2008}), we show in what sense  quantum vacuum energy
influences gravitational field, and in what sense it does not.
Actually, we propose a reasonable derivation of a contribution of
quantum vacuum energy which influences gravitational field, and
which assumes an experimentally reasonable value.

\section{Quantum vacuum energy}
The standard approach (Weinberg \cite{Wein1989}) to estimate the
quantum vacuum energy density $\varrho_{\rm vac}$ in the spirit of
Casimir energy yields for a single bosonic scalar mode
\begin{align}
  \varrho_{\rm vac}=\frac{1}{2}\int\limits_{0}^{\Lambda_{\rm \textsc{uv}}}
  \frac{4\pi c}{(2\pi\hbar)^3}\;
  \sqrt{(mc)^2+k^2}\;k^2\mathrm{d}k,
  \label{eq:wrong density 1}
\end{align}
where $m$ is the mass of the mode. For the planckian UV cutoff,
$\Lambda_{\rm \textsc{uv}}=\Lambda_{\rm \textsc{p}}=\sqrt{\hbar
c^3/G}\approx 6.5\rm\,kg\,m/s$, we obtain $\varrho_{\rm
vac}\approx 3.1\cdot 10^{111}\rm\, J/m^3$, whereas the
experimentally estimated value is of the order of the critical
density of the Universe, $\varrho_{\rm crit}={3
{\left(H_{0}c\right)}^{2}}/{8\pi G}\; (\approx 10^{-9} \rm\,
J/m^3)$.

In our opinion, the explicitly absurd result follows from an
erroneous approach. Namely, in classical as well as in quantum
theory interactions are being mediated by fields or particles. In
Eq.\,\eqref{eq:wrong density 1} no explicit or implicit coupling
to gravitational field appears on any stage. Therefore, by
construction, we assume that gravitation does not couple (is
insensitive) to the term \eqref{eq:wrong density 1}. As there is
no any coupling to \eqref{eq:wrong density 1}, its huge value is
isolated from the outer world and therefore invisible
(non-existent). What we have just said is, so to say, a negative
part of our reasoning. In the positive part we should cure the
situation somehow proposing a reasonable solution. In (Broda {\em
et al.\/} \cite{Broda2008}) we have sketched our idea and proposed
an estimation of the quantum vacuum energy. Actually, it is
possible to allow another interpretation of our calculations. For
example, in our opinion, the idea of ``the rearrangement'' of
vacuum motivated by thermodynamics and condensed-matter physics
advocated in (Volovik (\cite{Vol2005}, \cite{Vol2006})) could be
implemented just this way.

Anyway, our original calculus (Broda {\em et al.\/}
\cite{Broda2008}) consists in careful considering only
contributions coming from attached classical external lines. More
precisely, in the first step, we should estimate quantum vacuum
fluctuations of a matter field in the background of an external
classical gravitational field. In the next step we should retain
the most divergent part and subtract the term without
gravitational field.

\section{The estimation}
We will work in the formalism of the effective action, throughout.
A euclidean version of our approach has been given in (Broda {\em
et al.\/} \cite{Broda2008}), and here we present a relativistic
one. Full quantum contribution to the effective action coming from
a single (non-self-interacting) mode is (DeWitt
(\cite{DeWitt1975}, \cite{DeWitt2003}))
\begin{align}
 S_{\rm eff}=\pm\frac{i\hbar}{2}\log\det\mathcal{D},
 \label{eq:EffectiveAction1}
\end{align}
where $\mathcal{D}$ is a second-order differential operator, in
general, with classical external fields, and the upper (plus) sign
corresponds to a boson, whereas the lower (minus) one corresponds
to a fermion, respectively. Proper-time UV regularized version of
\eqref{eq:EffectiveAction1} in Schwinger's formalism assumes the
form (Birrell \& Davies \cite{Birrell1982}), (DeWitt
(\cite{DeWitt1975}, \cite{DeWitt2003}))
\begin{align}
  S^{\varepsilon}_{\rm eff}=\mp\frac{i\hbar}{2}\int\limits_{\varepsilon}^{\infty}\frac{i\mathrm{d}s}{is}\;\mathrm{Tr}
  \;e^{-is\mathcal{D}}.
 \label{eq:EffectiveActionRegularized}
\end{align}
Next, we apply the Seeley--DeWitt ``heat-kernel'' expansion in
four dimensions (Birrell \& Davies \cite{Birrell1982}), (DeWitt
(\cite{DeWitt1975}, \cite{DeWitt2003})), (Ball \cite{Ball1989}),
\begin{align}
\left<x\right|e^{-is\mathcal{D}}\left|x\right>=i(4
\pi)^{-2}\sum\limits_{n=0}^{\infty}a_{n}(x)(is)^{n-2}.
 \label{eq:Seeley--DeWitt Expansion}
\end{align}
The contribution coming from the first term, we are interested,
i.e.\ $a_{0}(x)$, is
\begin{align}
S_{\rm vac}=\mp\frac{\hbar}{2} \frac{1}{2 \varepsilon^2}
\frac{1}{\left(4 \pi\right)^2}\;\mathrm{Tr}\;a_{0}(x).
\label{eq:a0VacuumActionContribution 1}
\end{align}
Since $a_{0}(x)=1$, and for planckian UV cutoff
$\varepsilon=\frac{\hbar G}{c^3}$, we obtain
\begin{align}
S_{\rm vac}=\mp\frac{1}{4} \frac{c^7}{\left(4 \pi\right)^2 \hbar
G^2}\;\int \sqrt{-g}\,{\mathrm{d}}^3x\mathrm{d}t.
\label{eq:a0VacuumActionContribution2}
\end{align}
For simplicity, we confine ourselves to the spatially flat
Friedmann--Lema\^{\i}tre--Robertson--Walker metric with the scale
factor $a(t)$. To ease our calculus further, we set the present
coordinate time $t=0$, and normalize the coordinates to unity,
i.e.\ $a(0)=1$. Expanding $a(t)$ around $t=0$ we have
\begin{align}
a(t)=1+H_0t-\frac{1}{2}q_0{H_0}^2t^2+\mathcal{O}(t^3),
\label{eq:a(t)FactorExpansion1}
\end{align}
where $H_0$ is the present day Hubble expansion rate, and $q_0$ is
the present day deceleration parameter. Hence
\begin{align}
\sqrt{-g}=\left[a^2(t)\right]^{3/2}=\left[1+2H_0t+\left(1-q_0\right){H_0}^2t^2+\mathcal{O}(t^3)\right]^{3/2}.
\label{eq:DeterminantExpansion1}
\end{align}
Now, one can easily show that the infinitesimal gauge
transformation of the metric,
\begin{align}
\delta g_{\mu\nu}=\partial_{\mu}\xi_{\nu}+\partial_{\nu}\xi_{\mu},
\label{eq:Gauge transformation 1}
\end{align}
with the gauge parameter
\begin{align}
\xi_{\mu}=\left(\frac{1}{2}H_{0}{\bold{x}}^2, -H_{0}tx^{i}\right),
\label{eq:GaugeParameter}
\end{align}
cancels the linear in $t$ part in
\eqref{eq:DeterminantExpansion1}. There are also general arguments
supporting this cancellation given in (Shapiro
\cite{Shapiro2007}). Therefore
\begin{align}
\sqrt{-g}\approx 1+\frac{3}{2}\left(1-q_0\right){H_0}^2t^2,
\label{eq:DeterminantExpansion2}
\end{align}
and
\begin{align}
S_{\rm vac}\approx \mp\frac{1}{4} \frac{c^7}{\left(4 \pi\right)^2
\hbar G^2}\;\int
\left[1+\frac{3}{2}\left(1-q_0\right){H_0}^2t^2\right]\mathrm{d}t
\int{\mathrm{d}}^3x. \label{eq:a0VacuumActionContribution3}
\end{align}
The number one in the bracket corresponds to the term uncoupled to
gravitational field, and it should be subtracted. By the way, such
a subtraction is a standard procedure in quantum field theory. As
we are interested in a density rather than in a total value we
should get rid off all integrals. Since the integrand is only
time-dependent we can simply discard the spatial volume
$\int{\mathrm{d}}^3x$. As far as the time integrand is concerned
we should take into account that our calculus is perturbative in
$t$ and valid only in the vicinity of $t=0$. Therefore, we have to
take the limit of ``infinitesimal'' time. From the point of view
of quantum field theory the ``infinitesimal'' time is the Planck
time $T_{\rm \textsc{p}}=\sqrt{{\hbar G}/{c^5}}$. So, our density
is a time average, i.e.\ ${T_{\rm
\textsc{p}}}^{-1}\int\limits_{0}^{T_{\rm
\textsc{p}}}\mathrm{d}t(\cdot)$, and assumes the form
\begin{align}
\varrho\approx \mp\frac{1}{4} \frac{c^7}{\left(4 \pi\right)^2
\hbar G^2}\frac{1}{2}\left(1-q_0\right){H_0}^2{T_{\rm
\textsc{p}}}^2,
\label{eq:Final Density 1}
\end{align}
or finally
\begin{align}
\varrho\approx \mp\frac{1}{48 \pi}\left(1-q_0\right)\varrho_{\rm
crit},
\label{eq:Final Density 2}
\end{align}
where we have used the relation: ${H_0}^2=\frac{8}{3}\pi
\frac{G}{c^2}\varrho_{\rm crit}$. For, e.g.\ $q_0=-0.7$ (Virey
{\em et al.\/} \cite{Virey2005}), we get
\begin{align}
\varrho\approx \mp0.01\varrho_{\rm crit},
\label{eq:Final Density
3}
\end{align}
a very promising result.

\section{Conclusions}
In the framework of standard quantum field theory, without any
additional more or less exotic assumptions we are able to derive
an experimentally reasonable result \eqref{eq:Final Density 3}.
This numeric value corresponds to only a single mode. Therefore in
the real world it should be multiplied by a small natural number.

\section*{Acknowledgments}
This work was supported in part by the Polish Ministry of Science
and Higher Education Grant PBZ/MIN/008/P03/2003 and by the
University of  {\L}\'od\'z grant. One of the authors (B.B.) would
like to thank the organizers for their kind invitation and for
their generous support.


\end{document}